\documentclass[aps,superscriptaddress,long,notitlepage,letterpaper,balancelastpage,nofootinbib,prd,floatfix,twocolumn]{revtex4-1}
\pdfoutput=1
\usepackage{amsmath,mathtools,amssymb,amsthm,amsxtra,overpic,bbm,epsfig,url,bm}
\usepackage{hyperref}
\usepackage{mathrsfs}
\usepackage{color,xcolor}
\usepackage{comment}
\usepackage{float}
\usepackage{enumitem}
\usepackage{lmodern}
\usepackage{slashed}
\usepackage{multirow}
\usepackage{appendix}
\usepackage{tabularx}
\usepackage[T1]{fontenc}
\usepackage{caption}
\usepackage{subcaption}
\usepackage{graphicx}
\usepackage{slashed}
\usepackage{ragged2e}

\DeclareUnicodeCharacter{03BC}{\ensuremath{\mu}}
\DeclareUnicodeCharacter{2032}{\ensuremath{'}}
\definecolor{nicered}{rgb}{0.5,0.,0.}
\definecolor{nicegreen}{rgb}{0.,0.5,0.}
\definecolor{niceblue}{rgb}{0.,0.,0.5}
\hypersetup{
	colorlinks=true,
	linkcolor=black,
	filecolor=nicegreen,      
	urlcolor=niceblue,
	citecolor=nicered,
}

\newcommand{\prlsection}[2]{{\it\textbf{#1}{#2}}---}
\makeatletter
\newcommand*{\balancecolsandclearpage}{%
	\close@column@grid
	\cleardoublepage
	\twocolumngrid
}

\newcommand{\0}{$0\nu\beta\beta$}

\newcommand{\LnuEM}{L$\nu$EM}

\newcommand{\nubar}{\overline{\nu}}

\newcolumntype{Y}{>{\centering\arraybackslash}X}

\setlist{nolistsep}

\makeatother
\begin{document}

\title{\vspace{1cm} \Large 
Neutrinoless Double Beta Decay  without \\ Vacuum Majorana Neutrino Mass
}

\author{\bf Lukáš Gráf}
\email[E-mail: ]{lukas.graf@berkeley.edu}
\affiliation{Department of Physics, University of California, Berkeley, CA 94720, USA}
\affiliation{Department of Physics, University of California, San Diego, La Jolla, CA 92093-0319, USA}

\author{\bf Sudip Jana}
\email[E-mail: ]{sudip.jana@mpi-hd.mpg.de}
\affiliation{Max-Planck-Institut f{\"u}r Kernphysik, Saupfercheckweg 1, 69117 Heidelberg, Germany}

\author{\bf Oliver Scholer}
\email[E-mail: ]{scholer@mpi-hd.mpg.de}
\affiliation{Max-Planck-Institut f{\"u}r Kernphysik, Saupfercheckweg 1, 69117 Heidelberg, Germany}

\author{\bf Nele Volmer}
\email[E-mail: ]{volmer@mpi-hd.mpg.de}
\affiliation{Max-Planck-Institut f{\"u}r Kernphysik, Saupfercheckweg 1, 69117 Heidelberg, Germany}

\begin{abstract}
We present a proof-of-concept extension to the Standard Model that can generate a non-vanishing neutrinoless double beta decay (\0) signal without the existence of Majorana neutrinos or lepton number violation in the zero-density vacuum-ground-state Lagrangian. We propose that the \0 can be induced by the capture of an ultralight scalar field, a potential dark matter candidate, that carries two units of lepton number. This makes the observable \0 spectrum indistinguishable from the usual \0 mechanisms by any practical means. We find that a non-zero \0 rate does not require neutrinos to be fundamentally Majorana particles. However, for sizeable decay rates within the range of next-generation experiments, the neutrino will, generally, acquire an (effective) Majorana mass as the scalar field undergoes a transition to the Bose-Einstein condensate phase. We also discuss the distinction between the aforementioned scenario and the case in which the emission of a lepton-number-two scalar leads to \0, exhibiting discernible qualitative features that make it experimentally distinguishable from the conventional scenario.
\noindent 
\end{abstract}
\maketitle
\prlsection{Introduction}{.}
Ettore Majorana's formulation of the novel theory, asserting the indistinguishability of neutrinos and antineutrinos~\cite{Majorana:1937vz}, prompted Giulio Racah to propose a test for Majorana's theory~\cite{Racah:1937qq}, while in 1939, Wolfgang Furry was the first to contemplate neutrinoless double beta decay (\0)~\cite{Furry:1939qr}. Since the early stages of geochemical, radiochemical, and counter experiments, significant focus has been devoted to double-beta decay~\cite{Barabash:2011mf}. Over the past three decades, evidence from solar, reactor, atmospheric, and accelerator neutrino experiments~\cite{ParticleDataGroup:2022pth} has indicated non-zero neutrino masses and mixing. However, several fundamental questions about neutrinos still persist, particularly regarding their nature---whether they are Dirac or Majorana---and how to differentiate between them. In this context, \0 experiments~\cite{Umehara:2008ru, GERDA:2020xhi, CUPID-0:2018rcs, NEMO-3:2009fxe, CUPID:2020aow, Danevich:2016eot, Arnaboldi:2002te, CUORE:2019yfd, EXO-200:2017vqi, KamLAND-Zen:2022tow, KamLAND-Zen:2024eml, NEMO:2008kpp, Armengaud:2019loe, 2017LEGEND200,legendcollaboration2021legend1000, nEXO:2021ujk, SNO:2021xpa} offer the most sensitive laboratory probe not only for the Majorana nature of neutrinos but also for various lepton-number-violating scenarios~\cite{Cirigliano:2017djv, Graf:2018ozy, Cirigliano:2018yza, Fridell:2023rtr, Graf:2022lhj, Scholer:2023bnn}. For recent reviews of \0, see Refs.~\cite{2019review_Werner, Agostini:2022zub, Cirigliano:2022oqy}. It is widely acknowledged in the literature that the observation of \0 would conclusively establish neutrinos as Majorana particles~\cite{Schechter:1981bd, Takasugi1984372}. While the lepton number violating process causing \0-decay is known to induce a Majorana mass for the neutrino~\cite{Schechter:1981bd}, it may not necessarily be the dominant mechanism for neutrino mass generation~\cite{Duerr:2011zd}. In the literature, there are various new physics models that enhance the \0 rate and elucidate links between lepton number violation, double beta decay, and neutrino mass. For instance, see Refs.~\cite{Deppisch:2012nb, Rodejohann:2011mu}.
\begin{figure}[tb!]
\centering
  \includegraphics[width=\columnwidth]{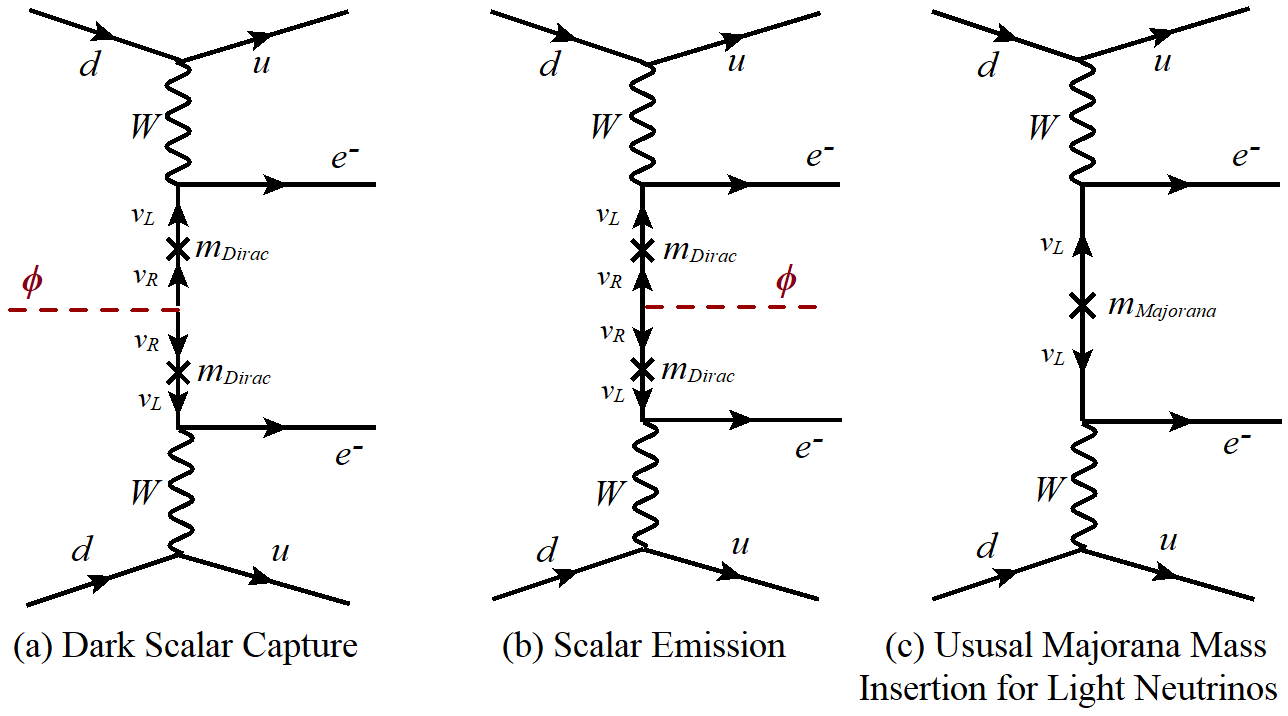}
  \caption{\justifying Representative Feynman diagrams for \0 with a with lepton-number-carrying scalar $\phi$: (a) the scalar capture and (b) the scalar emission diagram. Additionally, we show the usual \0 diagram (c) induced by the exchange of light Majorana neutrinos for comparison.}
\label{fig:feynman_diagrams}
\end{figure}
In contrast, here we propose a scenario that can generate non-zero \0 rates in a Dirac neutrino model while having an experimental signature that is indistinguishable from the standard \0 scenarios, which implies that it can generate a non-vanishing \0 signal without the existence of Majorana neutrinos or lepton number violation in the vacuum-ground-state Lagrangian. The motivation behind this proposition stems from the fact that if the lepton number, or any non-anomalous symmetry encompassing lepton number (such as $B-L$), is a conserved symmetry in nature, then neutrinos inherently exhibit Dirac particle characteristics\footnote{However, in scenarios where neutrinos are considered Dirac particles with intrinsically small masses, there exists a potential for them to function as pseudo-Dirac particles. Quantum gravity corrections, governed by higher dimensional operators suppressed by the Planck scale, could induce extremely small Majorana masses for neutrinos~\cite{Wolfenstein:1981kw, Petcov:1982ya, Valle:1983dk, Kobayashi:2000md}. These corrections adhere to all gauge symmetries within the theory; nonetheless, they are not anticipated to respect global symmetries, such as lepton numbers. For models addressing Dirac neutrino masses, see Refs.~\cite{Babu:2022ikf, Jana:2019mgj, Farzan:2012sa, Ma:2016mwh, Jana:2019mez, Bonilla:2018ynb, Jana:2021tlx, CentellesChulia:2019xky} and therein.}. To accommodate this, we introduce a scalar particle $\phi$ with a lepton number of two units,  which establishes interactions with the Standard Model (SM) neutrinos after the electroweak symmetry breaking. Here, we regard $B-L$ as an exact symmetry, and we structure the scalar potential in such a way that the $\phi$ field avoids obtaining a vacuum expectation value (vev). Consequently, this prevents the emergence of a vacuum Majorana mass term for light neutrinos. The coupling of $\phi$ bears resemblance to that of a Majoron, the Goldstone boson emerging from the spontaneously broken lepton number symmetry. If such symmetry breaking is accountable for neutrino mass generation,
the Majoron coupling is determined by $i m_\nu / \Lambda$, where $\Lambda$ signifies the symmetry breaking scale. On the contrary, it is crucial to note that the neutrino couplings to $\phi$ presented here are unrelated to or restricted by the vacuum neutrino masses. The phenomenology of such a leptonic scalar $\phi$ has been investigated in different contexts~\cite{Berryman:2018ogk, deGouvea:2019qaz, Kelly:2019wow}. Here, we propose such a simple scenario that can generate non-zero \0 rates in a Dirac neutrino model with the leptonic scalar $\phi$ while having an experimental signature that is indistinguishable from the usual \0 scenarios (cf. Fig.~\ref{fig:feynman_diagrams}). As we also discuss in detail, for large enough number densities of the scalar, an in-medium Majorana neutrino mass gets induced, which triggers a mechanism equivalent to the usual light-Majorana-neutrino-exchange mechanism (\LnuEM).

\prlsection{$\mathbf{0\nu\beta\beta}$ and Nature of Neutrinos}{.}\label{sec:general_discussion}
The black-box theorem, as formulated by Schechter and Valle~\cite{Schechter:1981bd}, posits that the presence of \0 would directly imply that neutrinos acquire a Majorana mass, notwithstanding their known minuteness~\cite{Duerr:2011zd}. In this context, our aim is to scrutinize potential scenarios with experimental signatures resembling usual \0 mechanisms (cf. Fig.~\ref{fig:feynman_diagrams}) while concurrently circumventing the implications presented by the black-box theorem.
\begin{figure}[t!]
	\centering
	\includegraphics[width=0.95\columnwidth]{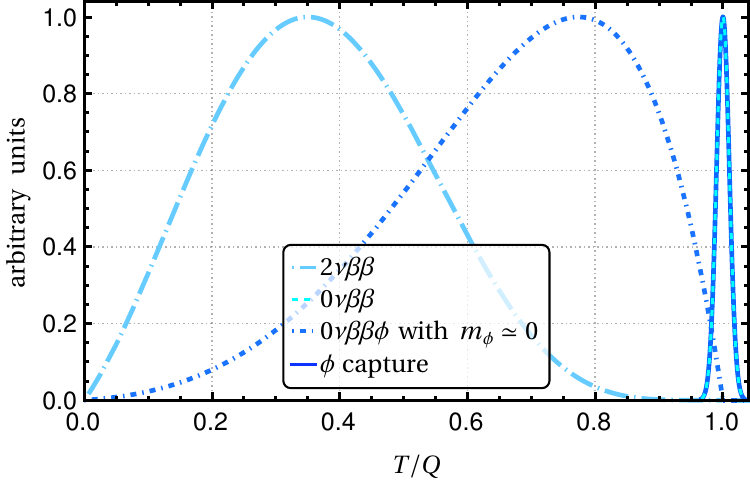}
	\caption{\justifying Total electron energy spectra for different double beta decay modes. On the x-axis, we normalized the summed kinetic electron energy $T=\epsilon_1+\epsilon_2-2m_e$ with respect to the $Q$ value, where the $\epsilon_i$ are the energies of the outgoing electrons, and $m_e$ is the electron mass. The dashed-dotted lines depict the standard double beta decay spectrum (lighter, long dash) and the distribution for \0 with emission of a scalar $\phi$ with mass $m_{\phi}\simeq 0$ (darker, short dash)~\cite{Brune_2019}. The spectra of both the \0 and \0 induced via capture of a scalar $\phi$ have the form of overlapping delta distributions smeared out by the experimental accuracy (exaggerated here for better visibility).
 }
	\label{fig:spectra}
\end{figure}
The black-box theorem generally holds for every transition of the form
\begin{align}
    (A,Z) \rightarrow (A, Z\pm 2) +2 e^\mp\label{eq:neutrinoless_transition_standard},
\end{align}
independently of the underlying particle physics. However, its applicability is limited in scenarios involving additional external particles, as noted in the literature, for instance, in the emission of a lepton number-carrying \textit{Majoron} (denoted by $\chi$)~\cite{Burgess:1992dt, Burgess:1993xh, Carone:1993jv}
\begin{align}
    (A,Z) \rightarrow (A, Z\pm 2) +2 e^\mp + n\chi,\label{eq:neutrinoless_transition_majoron_emission}
\end{align}
where $n$ signifies the quantity of emitted Majorons. While these models can instigate \0, their experimental signature differs from the usual \0 scenarios. 

In \0 experiments, the detection method involves measuring the collective energy of emitted electrons. While the total electron energy spectrum of the standard double beta decay is a continuous function between $0$ and the $Q$ value, in the neutrinoless mode represented by Eq.~\eqref{eq:neutrinoless_transition_standard}, all the decay energy is deposited in the emitted electrons. Notably, \0 variants with additional particle emissions result in a distinct energy spectrum, modifying the standard double beta decay spectrum as these particles can carry away some of the released decay energy.

Instead of the emission of a particle carrying a lepton number, one can conceive of \0 being triggered by the capture of one or more such particles
\begin{align}
    n\chi + (A,Z) \rightarrow (A,Z\pm2) + 2e^\mp.
    \label{eq:neutrinoless_transition_capture}
\end{align}
Fig.~\ref{fig:spectra} depicts a comparison of the energy spectra for the standard double beta decay, \0 with a scalar emission, the usual \0 as it is induced via, e.g., the \LnuEM, and the \0 induced by the capture of a scalar. If the energy transferred by the captured particle(s) remains below the experimental energy resolution, both the \0 and capture modes exhibit identical experimental signatures. To be more precise, in the model we present hereafter the energy spectra as well as the angular correlation of the outgoing electrons are expected to be the same in both scenarios. In the following, we will discuss \0 induced via the capture of a dark scalar from the cosmic background. The alternative possibilities of fermionic or vector-boson captures are briefly discussed in the \hyperref[appendix]{Appendix}.


\prlsection{$\mathbf{0\nu\beta\beta}$ Induced via a Dark Scalar Capture}{.}
Before delving into the analysis of \0 induced by dark scalar capture, we introduce a benchmark model for such a scenario. Alongside the SM particle spectrum, we add a scalar field $\phi$, a singlet under SM symmetries but charged under a global $B-L$ symmetry (with a lepton number of two units and $B-L$ charge of $-2$). Additionally, the introduction of the right-handed neutrino $\nu_R$ (with a $B-L$ charge of $-1$) is necessary for generating the Dirac neutrino mass term. The relevant part of the Lagrangian is as follows:
\begin{align} \label{eq:Lagrangian}
    \mathcal{L} \supset&  m_\phi^2 \phi^\dagger \phi + \lambda_\phi (\phi^\dagger \phi)^2 + \lambda_{H\phi}(H^\dagger H)(\phi^\dagger \phi) \nonumber
\\
    &+Y_{ij}^\nu \overline{L}_i \Tilde{H} \nu_{R,j} + g_{ij} \nubar_{R,i} \nu_{R,j}^C \phi +\mathrm{h.c.}
\end{align}
Enforcing $B-L$ as an exact symmetry prevents the emergence of Majorana neutrino mass. 
After electroweak symmetry breaking, neutrinos acquire a Dirac mass $m_{Dirac} = Y^\nu v_{EW}$, where $v_{EW}$ denotes the electroweak vev. By selecting suitable parameters for the scalar potential, the scalar field avoids acquiring a vev. Also, we require the Higgs portal coupling $\lambda_{H\phi}$ to be small. The absence of spontaneous $B-L$ breaking in the vacuum results in neutrinos exhibiting characteristics of Dirac particles. While one could argue that global symmetries should not be conserved in a theory of quantum gravity~\cite{Kallosh:1995hi}, it is straightforward to assume a gauged $U(1)_{B-L}$~\cite{Heeck:2014zfa}, which is then preserved even at the non-perturbative level. 

At first sight, there are two different processes that contribute to \0 within the proposed model. These are the capture of the scalar on a nucleus with a subsequent nuclear decay, and the \0 accompanied by the emission of $\phi$ similar to the Majoron emission scenarios
\begin{align}
     \phi + (A,Z) \longrightarrow (A,Z+2) + 2e^-,&\quad  (\text{scalar capture})\nonumber
     \\
     (A,Z) \longrightarrow (A,Z+2) + 2e^- + \phi^\dagger,& \quad (\text{scalar emission}).
\end{align}
The corresponding Feynman diagrams are shown in Fig.~\ref{fig:feynman_diagrams} (a) and (b). Here, we want to show that for ultralight scalars $m_\phi\ll 1\,\mathrm{eV}$ the capture in Fig.~\ref{fig:feynman_diagrams} (a) can generate sizeable decay rates in reach of next-generation \0 experiments while simultaneously suppressing the scalar emission diagram (Fig.~\ref{fig:feynman_diagrams} (b)) which would yield a different electron energy spectrum than the standard \LnuEM\ in Fig.~\ref{fig:feynman_diagrams} (c). Such ultralight scalars can be produced non-relativistically in the early universe (see, e.g.,~\cite{Ferreira:2020fam} and references therein).

Although not essential for our discussion, it is noteworthy that the scalar $\phi$ could potentially serve as cold dark matter (DM)~\cite{Sin:1992bg,Ji:1994xh,Hu:2000ke,Ferreira:2020fam}, depending on its mass, energy, and number density. If $\phi$ contributes to a fraction of the DM density, it must exhibit stability over cosmological timescales. This can be realized either by ensuring that $m_{\nu,\mathrm{min}}>m_\phi/2$ or by adjusting the coupling to the lightest neutrino accordingly. Relevant recent studies on \0 induced through interactions with dark matter are outlined in Refs.~\cite{Huang:2021kam, Nozzoli:2022tov, Herms:2023cyy}. In contrast to previous literature, we do not consider the presence of a vacuum Majorana mass term by explicitly assuming the conservation of $B-L$.


\prlsection{$\mathbf{0\nu\beta\beta}$ in the Free Phase}{.} We estimate the scalar capture rate $\Gamma^\mathrm{cap}_{0\nu\beta\beta\phi}$ and scalar emission decay rate $\Gamma^\mathrm{em}_{0\nu\beta\beta\phi}$ by comparison with existing calculations for \0 accompanied by Majoron emission~\cite{Hirsch:1995in,Kotila:2021mtq}. Assuming that the momentum carried by $\phi$ can be ignored in comparison to the typical Fermi momentum of nucleons, i.e., $p_\phi\lesssim\mathcal{O}(1\,\mathrm{MeV})\ll p_F\simeq\mathcal{O}(100\,\mathrm{MeV})$, the transition amplitude for the $\phi$ emission and capture processes is identical and reads
\begin{align}
    \begin{split}
        \mathcal{A}_{0\nu\beta\beta\phi} =& \frac{G_F^2}{2}\overline{e_L} (p_1) \gamma_\mu \sum_{ij} U_{ei}U_{ej} \frac{g_{ij}m_im_j}{q^4} \gamma_\nu e_L^C(p_2) 
        \\&\times J_{V-A}^\mu(1) J_{V-A}^\nu(2) + (p_1 \rightleftarrows p_2)
    \end{split}
\end{align}
where $U$ is the neutrino mixing matrix, $q$ is the momentum of the internal neutrino and $J_{V-A}$ is the standard weak charged current.
It is equivalent to the well-studied \LnuEM\ (see e.g.~\cite{Dekens:2020ttz}) except for the neutrino propagator, which in our model necessitates two additional neutrino mass insertions due to the coupling of $\phi$ to the right-handed neutrinos only. This transition amplitude can also be found in models of \0 accompanied by the emission of two scalars~\cite{Hirsch:1995in} and nuclear matrix elements (NMEs) have been obtained, recently, using the IBM2 model of nuclear structure~\cite{Kotila:2021mtq}. The corresponding phase-space factors (PSFs) are also available in the literature~\cite{Kotila:2015ata}. Specifically, for \textsuperscript{136}Xe we obtain $G_{0\nu\beta\beta\phi} = 4.09\times10^{-16}\,\mathrm{yr}^{-1}, G_{0\nu\beta\beta} = 1.458\times10^{-14}\,\mathrm{yr}^{-1},$~\cite{Kotila:2012zza} and $\mathcal{M}_{0\nu\beta\beta\phi} = 1.112\times10^{-3}$.
For simplicity, we assume a diagonal and real coupling, i.e., $g_{ij} = g\delta_{ij}$ ($g\in\mathbb{R}$) and we define the effective neutrino mass as
\begin{align}
    m_{\beta\beta\phi}^2 = \sum_i U_{ei}^2 m_i^2.
\end{align}
Assuming the lightest neutrino to be (almost) massless with normal mass ordering and vanishing CP phase we find $m_{\beta\beta\phi}^2 \simeq 78\,\mathrm{meV}^2$.
Consequently, we can write the decay rate of the scalar emission accompanied \0 as 
\begin{align}
    \Gamma^\mathrm{em}_{0\nu\beta\beta\phi} = g^2 \log(2) \left(\frac{m_{\beta\beta\phi}}{m_e}\right)^4 \left|\mathcal{M}_{0\nu\beta\beta\phi} \right|^2 G_{0\nu\beta\beta\phi}.
\end{align}
Notice that the decay rate of the scalar emission diagram is independent of the scalar number density. The $\phi$ capture rate is then given by
\begin{align}
    \Gamma^\mathrm{cap}_{0\nu\beta\beta\phi} = g^2 \log(2) \frac{\alpha \rho_\mathrm{DM}}{2m_\phi^2m_e^2} \left(\frac{m_{\beta\beta\phi}}{m_e}\right)^4 \left|\mathcal{M}_{0\nu\beta\beta\phi} \right|^2 G_{0\nu\beta\beta},
\end{align}
where $\rho_\mathrm{DM}\simeq 0.3\, \mathrm{GeV}/\mathrm{cm}^3$ is the local DM density~\cite{XENON:2018voc}, $\alpha$ is the fraction of the total DM mass that $\phi$ accounts for, $m_\phi$ is the mass of the scalar, and, for simplicity, we assumed $\phi$ to be non-relativistic today, i.e., $E_\phi\simeq m_\phi$. The scalar number density is then given by $n_\phi = \alpha \rho_\mathrm{DM}/m_\phi$.
Note that for the capture mode, the PSF corresponds to the PSF of the standard \LnuEM, as the final state is equivalent.
Comparing the emission and capture rates in \textsuperscript{136}Xe we find
\begin{align}
    \frac{\Gamma^\mathrm{em}_{0\nu\beta\beta\phi}}{\Gamma^\mathrm{cap}_{0\nu\beta\beta\phi}} = \frac{2m_\phi^2m_e^2}{\alpha\rho_\mathrm{DM}}\frac{G_{0\nu\beta\beta\phi}}{G_{0\nu\beta\beta}} \simeq \frac{6.4\times10^{-25}}{\alpha}\left(\frac{m_\phi}{10^{-20}\,\mathrm{eV}}\right)^2.
\end{align}
While the emission is largely suppressed for small scalar masses, the half-life corresponding to the $\phi$-capture rate in \textsuperscript{136}Xe reads
\begin{align}
    T^{1/2}_{0\nu\beta\beta\phi} = \frac{\log(2)}{\Gamma^\mathrm{cap}_{0\nu\beta\beta\phi}}\simeq \frac{1.4\times10^{28}\,\mathrm{yr}}{ \alpha g^2} \left(\frac{m_\phi}{10^{-20}\,\mathrm{eV}}\right)^2.\label{eq:free_half_life}
\end{align}
Note that the capture process is, in principle, distinguishable from the \LnuEM\ if the energy of the captured scalar exceeds the uncertainty of the energy measurement in the experiment. Then, the spectrum of the scalar capture process would be shifted beyond the $Q$ value given by $Q=E_i-E_f-2m_e$, where $E_{i,f}$ are the energies of the initial and final state nuclei. Additionally, as the NMEs differ from those of the usual \LnuEM, comparison of the half-lives in different isotopes could provide another possibility to disentangle the $\phi$ capture process~\cite{Graf:2022lhj}.

\prlsection{Transition to the Condensate Phase}{.} The realization of experimentally testable half-lives in current and future experiments necessitates large scalar number densities (low masses, cf. Eq.~\eqref{eq:free_half_life}). However, the above perturbative treatment of the scalar capture breaks down for such large scalar number densities corresponding to masses $m_\phi\ll1\,\mathrm{eV}$. In fact, $\phi$ will form a Bose-Einstein condensate (BEC)~\cite{Bose1924, Einstein1925} below a critical temperature~\cite{Yukalov:2011qj,Ferreira:2020fam}
\begin{align}
    T_C = \frac{2\pi}{m_\phi}\left(\frac{n_\phi}{\zeta(\frac{3}{2})}\right)^{2/3} \simeq 10^{30}\,\mathrm{eV}\times \alpha\left(\frac{10^{-20}\,\mathrm{eV}}{m_\phi}\right)^{(5/3)}.\label{eq:Tcrit}
\end{align}
For a given scalar temperature $T_\phi$, Eq.~\eqref{eq:Tcrit} can also be read in terms of a critical number density $n_{\phi,c}$ or, for a fixed mass density $\rho_\phi = \alpha \rho_\mathrm{DM}$, in terms of a critical scalar mass $m_{\phi,c}$ below which the field transitions to the BEC phase.
\begin{figure}[t]
	\centering
	\includegraphics[width=\columnwidth]{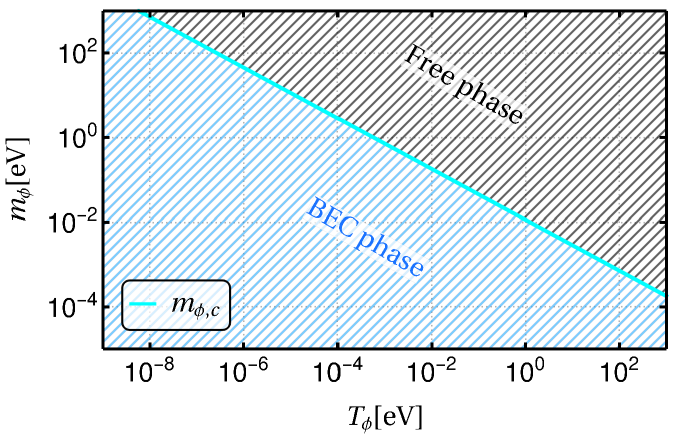}
	\caption{\justifying The transition from the free phase to the Bose-Einstein condensate phase is described in terms of the critical mass $m_{\phi,c}$ assuming a scalar mass density of $\rho_\phi = \rho_\mathrm{DM}$. Here, below $m_{\phi,c}$ the scalar is in the condensate phase and above it is in the free phase. The upper x- and y-axis range of the plot is set by the requirement that $m_\phi$ and $T_\phi$ remain below the typical energy resolution of \0 experiments of $\Delta E\sim\mathcal{O}(1\,\mathrm{keV})$~\cite{LEGEND:2021bnm}.}
	\label{fig:Tcritvsmphi}
\end{figure}
If the scalar field $\phi$ forms a BEC, its ground state acquires a non-trivial expectation value within the medium that spontaneously breaks the $B-L$ symmetry~\cite{Yukalov:2011qj,Laine:2016hma}. As a result, the right-handed neutrino receives an effective Majorana mass term via interactions with the medium. This effect resembles the consequences attributed to an effective refractive neutrino mass, as detailed in~\cite{Sen:2023uga}.

We can assess the validity of the perturbative treatment without considering the formation of the BEC by assuming that the critical temperature should not exceed the energy resolution of typical next-generation \0 experiments, such as LEGEND~\cite{LEGEND:2021bnm}, i.e., by requiring $T_C\lesssim1\, \mathrm{keV}$. This condition sets a limit on the scalar temperature and mass. The region of the $T_\phi-m_\phi$ parameter space corresponding to the free phase obtained under the requirement $T_\phi >T_C$ is identified in Fig.~\ref{fig:Tcritvsmphi}. 

\prlsection{$\mathbf{0\nu\beta\beta}$ in the Condensate Phase}{.} In the BEC phase, the ground-state of $\phi$ acquires a non-zero expectation value, and $\phi$ can be described in a mean-field description in terms of its ground state condensate $\phi_\mathrm{cond}$ and excitations $\phi_\mathrm{exc}$ as 
\begin{align}
    \phi = \frac{1}{\sqrt{2}}\left(\left<\phi_\mathrm{cond}\right> + \phi_\mathrm{exc}\right),
\end{align}
where $\left<\phi_\mathrm{cond}\right> = \sqrt{n_{\phi,\mathrm{cond}}/(2m_\phi)}$ is the expectation value for the ground state condensate in the limit of the ideal Bose gas~\cite{Laine:2016hma}. As we have seen previously, the large number density that is required to generate a considerable scalar capture rate results in a critical temperature much larger than, e.g., the temperature of the CMB. Therefore, we can safely assume $T_\phi\ll T_C$ such that $\phi$ is effectively completely described in terms of the BEC, i.e., $n_{\mathrm{cond}, \phi} \simeq n_\phi$ and $n_{\phi,\mathrm{exc}}\simeq 0$. In this context, within the scalar medium, the expectation value of the ground state condensate is equivalent to a vev generating a Majorana mass of the right-handed neutrino. 

Hence, for \0 it follows that the effects of the scalar condensate can be completely described in terms of the effective Majorana neutrino mass $m_{\beta\beta}$ for the electron neutrino, including the additional contribution from the three sterile neutrinos. The half-life can then be parameterized as~\cite{Bolton:2019pcu}
\begin{align}
        \left(T^{1/2}_{0\nu\beta\beta}\right)^{-1} = &\ g_A^4 G_{0\nu\beta\beta} \bigg|\sum_{i=1}^3 V_{ei}^2(m_\phi) \frac{m_{\nu,i}}{m_e} \mathcal{M}(0)\nonumber
       \\
        & + \sum_{i=4}^6 V_{ei}^2(m_\phi) \frac{m_{\nu,i}(n_\phi)}{m_e} \mathcal{M}(m_i)\bigg|^2,\label{eq:condensate_half_life}
\end{align}
where $V_{ei}(m_\phi)$ is the $6\times6$ neutrino mixing matrix, which depends on the heavy neutrino masses, and hence also on the scalar mass $m_{\phi}$ (or, equivalently, on number density $n_\phi$). Further, $m_e$ and $m_N$ are the mass of the electron and proton, respectively, and we used the simple interpolation for the mass-dependent NMEs,
\begin{align}
    \mathcal{M}(m_i) = \frac{4 m_N^2 |\mathcal{M}_\nu^{(9)}|}{\left<p^2\right> + m_i^2},\quad \left<p^2\right> = 4m_N^2\frac{|\mathcal{M}_\nu^{(9)}|}{|\mathcal{M}_\nu^{(3)}|}.
\end{align}
Details on the light and heavy neutrino-exchange NMEs $\mathcal{M}_\nu^{(3)}$ and $\mathcal{M}_\nu^{(9)}$, including the numerical values from Ref.~\cite{Deppisch:2020ztt}, are provided in the \hyperref[appendix]{Appendix}. The simple interpolation approach employed is sufficient for our discussion. However, a more sophisticated treatment can be found in~\cite{Dekens:2023iyc}.

\begin{figure}[t!]
    \centering
    \includegraphics[width=\columnwidth]{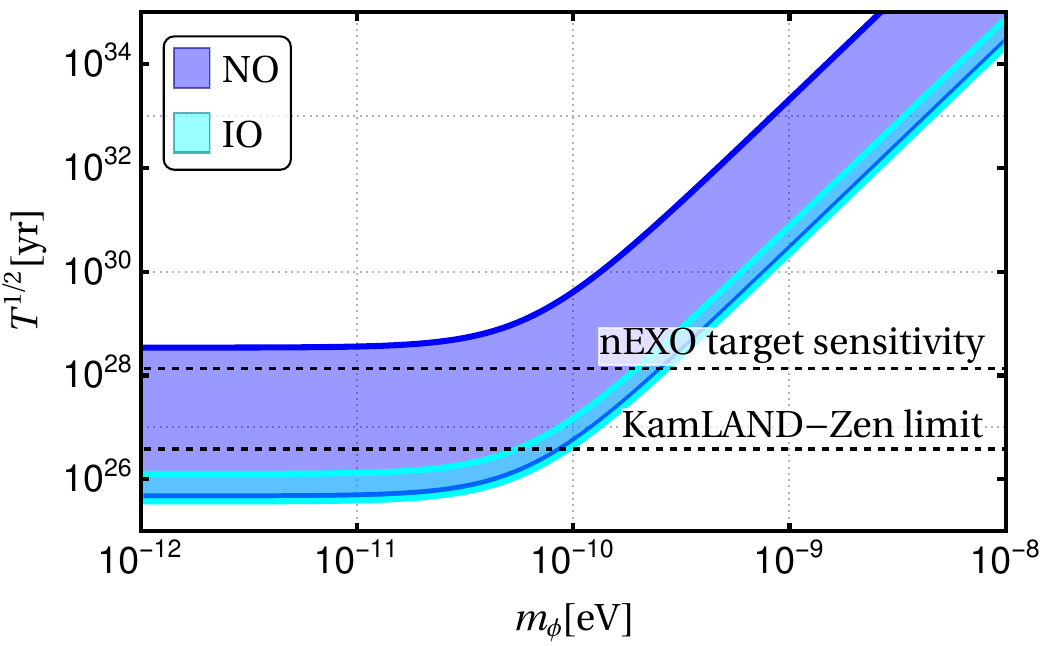}
    \caption{\justifying \0 half-life for \textsuperscript{136}Xe within the BEC phase in dependence on the scalar mass $m_\phi$. For larger scalar masses, i.e., smaller number densities, the neutrinos behave as pseudo-Dirac particles within the medium. Here, $g=\alpha=1$, the Dirac $CP$ phase is set to its best-fit value~\cite{ParticleDataGroup:2022pth} and the Majorana $CP$ phases are taken to be vanishing. The mass of the lightest neutrino is varied between zero (upper contours) and the highest possible value (lower contours), such that $\sum_{i=1}^3 m_i \le 260\,$meV~\cite{ParticleDataGroup:2022pth}. The blue region corresponds to normal ordering and the turquoise one to inverted ordering. For reference, we show the most recent half-life limit from KamLand-Zen~\cite{KamLAND-Zen:2024eml} and the projected target sensitivity of nEXO~\cite{nEXO:2021ujk}.}
    \label{fig:thalf}
\end{figure}

While the effective Majorana neutrino mass is, in principle, a function of the scalar density (or rather the expectation value), all the oscillation parameters are fixed from experiment such that, for a fixed scalar mass $m_\phi$, coupling $g$ and DM fraction $\alpha$ the only remaining free parameters are the minimal neutrino mass $m_\mathrm{min}$, the neutrino mass ordering, and the Majorana $CP$ phases. In Fig.~\ref{fig:thalf}, we show the dependence of the \0 half-life in \textsuperscript{136}Xe on the scalar mass $m_\phi$ for normal (NO) and inverted (IO) neutrino mass ordering with $g=\alpha=1$, the best-fit neutrino oscillation data from PDG~\cite{ParticleDataGroup:2022pth} and vanishing Majorana $CP$ phases. The minimal neutrino mass is varied within the allowed range by setting $\sum_{i=1}^{3} m_i\leq 260\,\mathrm{meV}$~\cite{ParticleDataGroup:2022pth}. It can be seen that for large scalar number densities (i.e., small masses) the usual half-life regime of the \LnuEM\ is reproduced, while for smaller number densities (i.e., larger masses) the neutrinos transition into a pseudo-Dirac regime. While such a pseudo-Dirac scenario can be strongly constrained from oscillation data as well as cosmology~\cite{Huang:2022wmz} it is not our main region of interest. In fact, for $m_\phi\lesssim 10^{-10}\,\mathrm{eV}$ (assuming $g=\alpha=1$), the active and sterile neutrino states are decoupled almost completely in our scenario. For a real scalar, time dependence of $\langle\phi_\mathrm{cond}\rangle \rightarrow \langle\phi_\mathrm{cond}\rangle \cos{(m_\phi t)}$ sets strong constraints on the coupling $g$ from oscillation data~\cite{Berlin:2016woy, Krnjaic:2017zlz}.
These limits do not extend to a complex scalar, as $\langle\phi_\mathrm{cond}\rangle\rightarrow\langle\phi_\mathrm{cond}\rangle\exp{(im_\phi t)}$ results in a constant $|\langle\phi_\mathrm{cond}\rangle|$.

\prlsection{Other Phenomenological Implications}{.}
While our extension to the SM is rather simple, it comes with a variety of new phenomenological features that should be taken into account. 
Numerous processes across various energy scales can detect the presence of the scalar $\phi$, which can be generated through radiation from a neutrino. High-energy collider experiments, for instance, may result in a distinctive same-sign dilepton signature accompanied by missing energy and forward jets~\cite{deGouvea:2019qaz}. This occurs as $\phi$ is generated in the vector-boson fusion process and subsequently manifests missing energy. Emitting the leptonic scalar $\phi$ from neutrino legs contributes to decay modes of various particles, including $Z$ boson ($Z \to \nu \nu \phi$), $W$-boson ($W \to \ell \nu \phi$), SM Higgs boson ($h \to \nu \nu \phi$), tau lepton ($\tau \to e \nu \nu \phi, \mu \nu \nu \phi$), muon ($\mu \to e \nu \nu \phi$), and charged mesons ($P \rightarrow \ell \nu \phi$ with $P=\pi, K, D, D_S, B$). In neutrino beam experiments, when neutrinos interact with matter fields, the emission of $\phi$ from the neutrino beams results in unique momentum distributions of charged leptons in the final state. Additionally, the outgoing charged lepton exhibits a wrong sign compared to standard scenarios, a feature easily detectable in magnetized detectors such as MINOS and DUNE~\cite{Berryman:2018ogk}. Despite stringent laboratory limits, our chosen mass and coupling combination relaxes these constraints due to the scalar's direct coupling to $\nu_R$ within our model, which in turn requires one or two neutrino mass insertions. 

The discussed model can have important implications in astrophysics and cosmology. The ultra-light scalar proposed could induce substantial neutrino self-interactions, impacting neutrinos from astrophysical sources traveling through the cosmic neutrino background. This may result in spectral distortions observable in experiments like IceCube~\cite{Ioka:2014kca, Ng:2014pca}. Additionally, these interactions may influence neutrino mixing patterns in astrophysical environments situated in regions characterized by varying dark matter densities. Constructing a renormalizable model for significant neutrino self-interactions is challenging. In our proposed model, the leptonic scalar $\phi$ offers large long-range neutrino self-interactions, along with short-range interactions within the condensate phase through the exchange of massive excitations~\cite{Ferreira:2020fam}. This has the potential to influence neutrino free-streaming during recombination and could address the Hubble tension as a byproduct~\cite{DiValentino:2021izs, Kreisch:2019yzn, Berbig:2020wve, Berryman:2022hds}. In our scenario, within the parameter region of interest, the sterile neutrinos are effectively decoupled from the active neutrinos such that limits from active-sterile oscillations or Big Bang nucleosynthesis are considerably weaker or do not apply. The exact interplay will be explored in more detail in future work. Last but not least, the hypothesis of a BEC dark matter has been studied intensively within the literature during the past decades, see e.g.,~\cite{Sin:1992bg, Ji:1994xh, Hu:2000ke, Boehmer:2007um, Das:2014agf, Sharma:2018ydn, Castellanos:2019ttq, Ferreira:2020fam, Antypas:2022asj}. For our scenario, observational limits drawn from gravitational effects of the scalar mass density $\rho_\phi$, such as structure formation, can be relaxed if $\phi$ is not assumed to explain all of the observed DM, i.e. if we choose $\alpha<1$.

In general, existing bounds on the coupling $g$ can be relaxed by increasing the number density $n_\phi$ as long as both parameters are not constrained simultaneously.

\prlsection{Conclusion}{.} 
The evidence for lepton number violation that would be implied by a future observation of neutrinoless double beta decay is usually automatically assumed to be tied to the Majorana nature of neutrinos, as it stems from the black-box theorem. The purpose of this work is to point out several subtleties that may arise in this connection for certain scenarios. Specifically, if an extra particle carrying lepton number takes part in \0, the entire process can preserve the $B-L$ symmetry. This is, for instance, the case for \0 with emission of a scalar field with $L=2$, but even more intriguingly, for \0 triggered by a hypothetical capture of such a scalar from a cosmic background. As discussed above, assuming a small momentum of the incoming scalar particle, this \0 mechanism would lead to a signature in principle indistinguishable from the usual mass mechanism. Although this observation is certainly interesting, the corresponding rate of the \0 induced by the scalar capture would be beyond the scope of any future experiment unless the number density of the considered scalar field is assumed to be large. In such a case, however, the scalar field forms a Bose-Einstein condensate, inducing an in-medium expectation value that effectively breaks the $B-L$ symmetry. While the $B-L$ number remains conserved in the vacuum ground state, neutrinos acquire an effective in-medium Majorana mass, which in turn triggers \0. In this way, the implication pointing from lepton number violation towards Majorana neutrino mass is restored for testable \0 half-lives, but the reasons for its validity, as pointed out, are quite distinct and more subtle than in usually considered scenarios.

\prlsection{Acknowledgments}{.}
The authors thank Thede de Boer, Vishnu P.K., Alexei Smirnov, and Andreas Trautner for helpful discussions. We are also grateful to Kaladi Babu, Frank Deppisch, Frederik Depta, George Fuller, Guo-yuan Huang, Werner Rodejohann, and Manibrata Sen for their useful comments on the final version of the manuscript.
LG acknowledges support from the National Science Foundation, Grant PHY-1630782, and to the Heising-Simons Foundation, Grant 2017-228.
\vspace{11pt}
\begin{center}
    \large{\textbf{Appendix}}\label{appendix}
    \label{app}
\end{center}
\vspace{11pt}

\prlsection{Capture of Dark Vector Bosons}{.}
In general, since vector bosons follow the same Bose-Einstein statistics, vector fields can form a BEC in the same sense as scalar fields can. However, due to the different Lorentz structures, their expectation value cannot result in a typical fermion mass term from a vector-interaction. Instead, neutrinos coupled to a vector background field can gain a refractive mass in terms of elastic forward scattering~\cite{Sen:2023uga}. Therefore, we expect no qualitative difference in the case of a vector boson capture induced \0. Moreover, the realization of such a model presents substantial challenges and complexity.

\prlsection{Capture of Dark Fermions}{.}\label{sec:fermion_capture}
Even within the SM, a \0 signature without the necessity for Majorana neutrinos is possible via the capture of cosmic relic neutrinos. This has been pointed out previously in, e.g.,~\cite{Hodak:2009zz,Hodak:2011zza} where the authors found a half-life of
\begin{align}
    \left(T_{\nu\nu\beta\beta}^{1/2}\right)^{-1} \simeq 7 \times10^{47}\,\mathrm{yr} \ \times \left(\frac{n_\nu}{\left<n_\nu\right>}\right)^2 
\end{align}
for a hypothetical ton-scale experiment using \textsuperscript{100}Mo and assuming an expected neutrino number density per neutrino type of $\left<n_\nu\right> = 56\,\mathrm{cm}^{-3}$. Similar mechanisms can arise in $R$-parity breaking supersymmetric extensions to the SM via the capture of DM neutralinos~\cite{Hirsch:1995ek,Hirsch:1995zi,Bolton:2021hje}. While a strong local relic neutrino clustering could, in principle, lead to observable signatures this clustering is fundamentally constrained by the Pauli exclusion principle. For a degenerate fermion gas with $p<p_\mathrm{max}$, the maximum number density is simply given by
\begin{align}
    n_\mathrm{max} = \frac{g_s}{(2\pi)^3} \frac{4}{3}\pi p_\mathrm{max}^3.
\end{align}
Here $g_s$ represents the number of spin degrees of freedom and we take $g_s = 2$ from here on. Hence, to put an upper limit on the number density, we need to put an upper limit on $p_\mathrm{max}$. For massless fermions, a reasonable upper bound is to assume that their energy density should not exceed the average photon energy density of the CMB of $\rho_\gamma \simeq 260\,\mathrm{meV}/\mathrm{cm}^{3}$~\cite{ParticleDataGroup:2022pth}. This gives $n_\mathrm{max}\simeq 4\times10^{3}\,\mathrm{cm}^{-3}$. 
For massive fermions, we can take the local dark matter density as an upper bound on the mass density and 
use the well-known Tremaine-Gunn bound~\cite{Tremaine:1979we} that puts a 
\begin{table}[h]
\caption{\justifying Summary of the relevant LECs taken from~\cite{Scholer:2023bnn}.}\label{Tab:LECs}
\center
\begin{tabularx}{\columnwidth}{ c Y  c Y }
\hline\hline
 \multicolumn{2}{c}{ $n\rightarrow pe\nu$, $\pi \rightarrow e \nu$ } &  \multicolumn{2}{c}{$\pi \pi \rightarrow e e$} \\
 \hline
 $g_A$ & $1.271$~\cite{ParticleDataGroup:2016lqr}      & $g^{\pi\pi}_{1}$   		& $  0.36 $~\cite{Nicholson:2018mwc}  \\
 \hline
  \multicolumn{2}{c}{$n \rightarrow p\pi ee$} & \multicolumn{2}{c}{$nn\rightarrow pp\, ee$}   
 \\
  \hline 
   $|g^{\pi N}_{1} |$       & $\mathcal{O}(1)$    & $|g^{N N}_1|$         & $\mathcal{O}(1)$
  \\ 
			   & 		          & $ g^{NN}_{\nu}$ & $ -92.9\,\mathrm{GeV}^{-2}$~\cite{Cirigliano:2020dmx,Cirigliano:2021qko,Wirth:2021pij}  
  \\\hline\hline
\end{tabularx}
\end{table}

\begin{table}[h]
\caption{\justifying Summary of the relevant NMEs obtained within the IBM2 framework~\cite{Deppisch:2020ztt} for \textsuperscript{136}Xe.}\label{tab:NMEs}
\center
\begin{tabularx}{\columnwidth}{YYYYY}
\hline\hline
\multicolumn{5}{c}{Long-Range}\\
\hline
 $M_F$ & \multicolumn{2}{c}{$M_{GT}$} & \multicolumn{2}{c}{$M_T$} 
 \\
 $-0.522$ & \multicolumn{2}{c}{$5.704$} & \multicolumn{2}{c}{$0.092$} 
 \\\hline
\multicolumn{5}{c}{Short-Range}\\
\hline
 $M_{F,sd}$ & $M_{GT,sd}^{AP}$ &  $M_{GT,sd}^{PP}$ & $M_{T,sd}^{AP}$ & $M_{T,sd}^{PP}$
 \\
 $-0.734$ & $-0.690$ & $0.167$ & $-0.363$ & $0.115$
 \\\hline\hline
\end{tabularx}
\end{table}
\noindent
lower limit of $m\gtrsim \mathcal{O}(100\,\mathrm{eV})$ on fermionic dark matter bound within galactic halos by requiring that $p_\mathrm{max}$ should not exceed the escape velocity of typical galaxies. Putting these two together results in an upper limit on the fermionic number density of $n_\mathrm{max}\simeq3\times 10^6\,\mathrm{cm}^{-3}$. While these bounds can be relaxed by assuming multiple species of fermions~\cite{Davoudiasl:2020uig}, the number of species required scales with $N_F\gtrsim \left(100\,\mathrm{eV}/m_f\right)^4$~\cite{Davoudiasl:2020uig}. Hence, local number densities on the order of $n_f/\left<n_\nu\right>\sim 10^{10}$ would require a significant amount of fermionic species $N_F\gtrsim 10^{21}$. It follows that for a fermionic capture model to reproduce observable half-lives, one would require either a very large number of fermionic DM species or a much larger interaction cross section than for the relic neutrino capture model. Additionally, one can expect single $\beta-$decay experiments like KATRIN~\cite{KATRIN:2005fny,KATRIN:2021uub,KATRIN:2022ayy} to be more sensitive to such scenarios~\cite{Hodak:2009zz,Hodak:2011zza}.

\prlsection{Nuclear Matrix Elements}{.}
For our calculations of \0 half-lives, we utilize the chiral EFT framework developed in~\cite{Cirigliano:2017djv,Cirigliano:2018yza} and automated in~\cite{Scholer:2023bnn}.
The long-range ($\mathcal{M}_\nu^{(3)}$) and short-range ($\mathcal{M}_\nu^{(9)}$) NMEs used in Eq.~\eqref{eq:condensate_half_life} are given as
\begin{align}
\begin{split}
    \mathcal{M}_\nu^{(3)} =& \frac{M_F}{g_A^2} - {M}_{GT} - {M}_{T} - 2\frac{m_\pi^2g_\nu^{NN}}{g_A^2} M_{F,sd}     
\end{split},
    \\
\begin{split}
    \mathcal{M}_\nu^{(9)} =& \frac{5}{6}\frac{g_1^{\pi\pi} m_\pi^2}{m_N^2}\bigg(\frac{1}{2}M_{GT,sd}^{AP}+M_{GT,sd}^{PP}
    \\
    &\qquad\quad\:\:\,+\frac{1}{2}M_{T,sd}^{AP}+M_{T,sd}^{PP}\bigg)
    \\
    &+\left(g_1^{\pi N}-\frac{5}{6}g_1^{\pi\pi}\right)\frac{m_\pi^2}{2m_N^2}\left(M_{GT,sd}^{AP}+M_{T,sd}^{AP}\right)
    \\
    &-2\frac{g_1^{NN}}{g_A^2}\frac{m_\pi^2}{m_N^2}M_{F,sd}.
\end{split}
\end{align}
Here, the $g_x^{yz}$ denote so-called low-energy constants (LECs) that represent the couplings that describe interactions in the chiral EFT Lagrangian and $m_\pi$ is the pion mass. For convenience, we have summarized the relevant LECs in Tab.~\ref{Tab:LECs}. The relevant NMEs obtained within the IBM2 framework for \textsuperscript{136}Xe are given in Tab.~\ref{tab:NMEs}. For the two unknown LECs $g_1^{\pi N},g_1^{NN}$ we use an estimate from naive dimensional analysis~\cite{Manohar:1983md} of $g_1^{\pi N}=g_1^{N N}=1$ in our calculation.

\bibliographystyle{utcaps_mod}
\bibliography{reference}
\end{document}